\documentclass[conference,10pt]{IEEEtran}
\pdfoutput=1

\usepackage{makeidx}
\usepackage{amsmath}
\usepackage{amssymb}
\usepackage{xcolor}
\let\labelindent\relax
\usepackage{enumitem}
\usepackage{graphicx}
\usepackage[backgroundcolor=yellow]{todonotes}
\usepackage{stfloats}

\newcommand{\tight}{\hspace{-0.4mm}}

\usepackage{subfig}
\usepackage{mathtools}

\let\originalleft\left
\let\originalright\right
\renewcommand{\left}{\mathopen{}\mathclose\bgroup\originalleft}
\renewcommand{\right}{\aftergroup\egroup\originalright}

\newcommand{\mref}[1]{\eqref{math:#1}}
\newcommand{\fref}[1]{Fig.~\ref{fig:#1}}

\newcommand{\geo}[2]{\frac{1-#1^{#2}}{1-#1}}
\newcommand{\geok}[2]{\frac{1-\left(#1\right)^{#2}}{1-#1}}

\input{setup_math.tex}

\usepackage{supertabular}
\usepackage[nonumberlist]{glossaries}
\makeglossaries
\glossarystyle{super}
\AtBeginDocument{%
	\abovedisplayskip 1.4ex plus4pt minus2pt%
	\belowdisplayskip \abovedisplayskip%
	\abovedisplayshortskip 0pt plus4pt%
	\belowdisplayshortskip 1.4ex plus4pt minus2pt
}

\begin{document}

\title{Analytical Model for IEEE 802.15.4 Multi-Hop Networks with Improved Handling of Acknowledgements and Retransmissions}

\author{\IEEEauthorblockN{Florian Meier}
\IEEEauthorblockA{Institute of Telematics\\
Hamburg University of Technology\\
Hamburg, Germany\\
florian.meier@tuhh.de}
\and
\IEEEauthorblockN{Volker Turau}
\IEEEauthorblockA{Institute of Telematics\\
Hamburg University of Technology\\
Hamburg, Germany\\
turau@tuhh.de}
}

\maketitle              
\thispagestyle{plain}
\pagestyle{plain}

\begin{abstract}
The IEEE 802.15.4 standard allows for the deployment of cost-effective and energy-efficient multi-hop networks. This document features an in-depth presentation of an analytical model for assessing the performance of such networks. It considers a generic, static topology with Poisson distributed data-collection as well as data-dissemination traffic. The unslotted CSMA/CA MAC layer of IEEE~802.15.4 is closely modeled as well as an enhanced model of the neighborhood allows for consideration of collisions of packets including interferences with acknowledgements. The hidden node problem is taken into account as well as a formerly disregarded effect of repeated collisions of retransmissions. The model has been shown to be suitable to estimate the capacity of large-scale multi-hop networks.
\end{abstract}

\makeatletter
\def\l@section#1#2{\addpenalty{\@secpenalty}\addvspace{0.0em plus 1pt}%
    \@tempdima 2.75em \begingroup \parindent \z@ \rightskip \@pnumwidth%
    \parfillskip-\@pnumwidth {\bfseries\leavevmode #1}\hfil\hbox to\@pnumwidth{\hss #2}\par%
    \endgroup}
\makeatother
\tableofcontents

\section{Introduction}
Analytical models have proven to be very suitable and effective for modeling large scale wireless mesh networks \cite{meiermodel}. This document is meant to present the model used in this paper in detail to allow for better understanding of the underlying mathematics and better reproduction of the results. As further reference, the reader might also be interested in the work by Di Marco et~al.~\cite{dimarco_analytical_2012} that served as a starting point to build this model. The following contributions were newly established in the model presented below:
\begin{itemize}
\item Downstream traffic from the central gateway to the nodes.
\item Collisions of acknowledgements with packets or other acknowledgements by enhancing the conflict graph.
\item Revised computation of collision probabilities after failures, taking into account simultaneous retransmissions.
\item Minor enhancements to allow for faster computation.
\end{itemize}

The source code of the implementation for reproduction and further use is published at GitHub \cite{sourceAnalyticalMultiHop}.

\begin{figure}[htb]
\centering
	\includegraphics[]{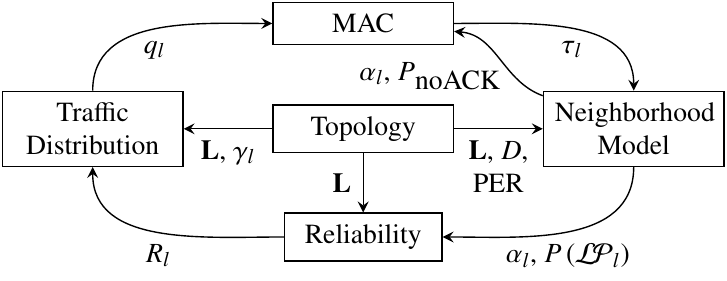}
\caption{Relationships between the different parts of the model.\label{fig:relations}}
\end{figure}

Fig. \ref{fig:relations} shows the different involved modules of the analytical model and how they are linked. The topology as presented in Sect. \ref{sec:topo} is given as an input to the model and describes how the nodes are linked and how they route the traffic. These information are used for calculating how the actual traffic is distributed within the network (Sect. \ref{sec:traffdist}) as well as for determining which transmissions might lead to busy channels and packet collisions in the neighborhood model (Sect. \ref{sec:neighbour}).

The traffic distribution yields the probability $\Psendl$ that a node has a pending transmission to send over link $l$. Though, it does not directly send the packet, but uses a CSMA/CA technique with retransmissions to increase the success probability. This is modeled in Sect.~\ref{sec:csma} and yields the probability $\t{l}$ that it actually senses the channel to start the transmission. For this, it also depends on information from the neighborhood model, such as the probability that the channel is sensed free $\a{l}$ and the probability that no acknowledgement arrives $\Pnoack$ and a retransmission should be started.

The information from the neighborhood model and the topology is also used to calculate the final reliability of a link $\R{l}$, that is the probability that a packet will eventually be transmitted over the link and not dropped because of a busy channel or repeated collisions. $\R{l}$ is again used to calculate the traffic distribution. This closes a circle, so it is evident that the equations of the outer modules are interlinked. They built up a non-linear equation system that can be computed numerically. For this task, the implementation utilizes the PETSc framework \cite{petsc-user-ref,petsc-efficient}. The topology, and thereby also the analog model, are not in the loop, so they can be calculated offline before the actual computation.

\section{Topology}
\begin{figure}[htb]
\centering
	\includegraphics[]{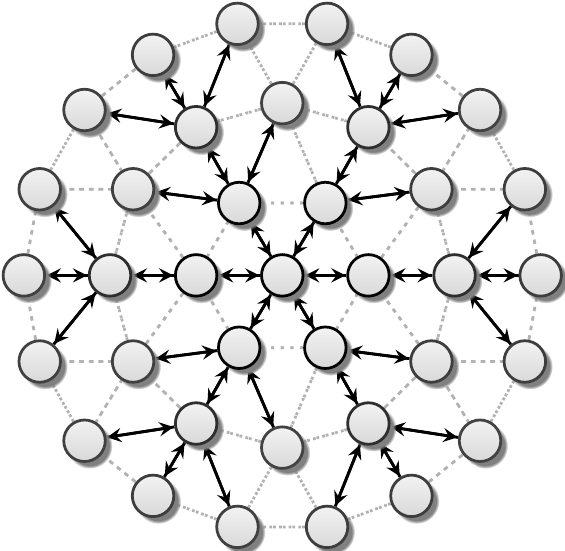}
\caption{A multi-hop network with a central gateway.\label{fig:network_bigger}}
\vspace{-0.2cm}
\end{figure}

The model considers a multi-hop network as depicted in Fig. \ref{fig:network_bigger}, with $\Nodes$ nodes including the gateway $\rootnode$ and $N-1$ clients. The dashed and continuous lines in Fig. \ref{fig:network_bigger} depict the predicate $\inrange{v,w}$ that a signal sent out by $v$ is strong enough to disturb an ongoing reception at $w$. This boolean predicate is calculated by an analog model together with the estimated bit error rate $\BER{(v,w)}$ of a data transmission between $v$ and~$w$.
\label{sec:topo}

An exemplary analog model for IEEE 802.15.4 networks that is used in the implementation is given in the following. It is presented in the IEEE 802.15.4 standard Annex~E~\cite{802154} and is based on a breakpoint log-distance path loss model determined by measurements. The received signal power in $\mbox{dB}$ with distance $d_{v,w}$ and transmission power $\Ptx$ is given by
\begin{align}
\Prx(v,w) &= \Ptx\tight-\tight\begin{cases}58.5 + 33\log_{10}\left(d_{v,w}/8\,m\right) &\hspace{-2mm} d_{v,w}\tight>\tight8\,m\\
40.2 + 20\log_{10}\left(d_{v,w}\right) &\hspace{-2mm} d_{v,w}\tight\leq\tight8\,m\end{cases}.
\end{align}
It also includes the computation of the bit error rate dependent on the noise power $\PN$
\begin{align}
\SNR{(v,w)} &=  10^{\frac{1}{10}\cdot\left(\Prx(v,w)-\PN\right)},\\
\BER{(v,w)} &= \frac{8}{15}\frac{1}{16}\sum_{k=2}^{16} -1^k \binom{16}{k} e^{20\cdot\SNR{(v,w)} \cdot\left(\frac{1}{k}-1\right)}
\intertext{and thereby the packet error rate for transmissions of $b$ bytes}
\PER{b,(v,w)} &= 1 - \left(1-\BER{l}\right)^{8 \cdot b}.\label{eq:per}
\end{align}
The predicate $\inrange{v,w}$ can be calculated from this, assuming there is a minimum interference power $\Pdist$ that is strong enough to disturb an ongoing reception at $w$
\begin{align}
    \inrange{v,w} \coloneqq \Prx(v,w) > \Pdist.
\end{align}

In a further preprocessing step, an optimal static routing tree \tree is computed by using Dijkstra's algorithm. The weight
\begin{align}
-\log\left(1 - \BER{(v,w)}\right) + 10^{-3}
\end{align}
was used in our experiments to minimize the bit error along the path, but still minimize the hop count if the BER is negligible. Though, any procedure that yields a suitable routing tree \tree can be used. The tree is described by the predicate $\T{p,c}$, that is true if and only if $p$ is a parent of $c$ in \tree, yielding the set of active links%
\begin{align}
\L \coloneqq \Lup &\cup \Ldown,\\
    \Lup \coloneqq \left\{(c,p)\,|\, \T{p,c} \right\},&\quad
    \Ldown \coloneqq \left\{(p,c)\,|\, \T{p,c} \right\}.
\end{align}

\section{Traffic Generation}
A Poisson traffic generation model is applied, so the intervals between two packet generations are exponentially distributed. The mean interval for traffic from a node to the gateway (upstream) is denoted as $\Iup$, while for each client, the gateway generates packets with mean interval $\Idown$. Analog to the IEEE 802.15.4 standard~\cite{802154}, all times are defined as multiples of
\begin{align}
\Sb\coloneqq\text{\stdname{aUnitBackoffPeriod}}\cdot\text{Symbol duration}\tight=\tight20\cdot 16\,\mu s.
\end{align}
Taking into account 4 bits per symbol for the O-QPSK PHY, a transmission of a packet of length $\BytesPacket$ bytes takes
\begin{align}
\Len = \frac{8\cdot \BytesPacket \cdot 16\,\mu s}{4 \cdot \Sb} = \frac{B}{10}
\end{align}
time units. An acknowledgment has a fixed length of $\BytesACK = 11$~bytes. Therefore it takes
\begin{align}
\Lack = \frac{11}{10}
\end{align}
time units. Together with the specified supplementary values $\IFS = 40$ and $\tack  = 12$, an acknowledged transmission takes
\begin{align}
\LenSuccess = \Len+\Lack+\frac{\IFS + \tack}{20} = \Len+\Lack+2.6
\end{align}
time units. The number of symbols $\tmACK$ to wait for a lost acknowledgement (\stdname{macAckWaitDuration}) is calculated according to the standard
\begin{align}
\begin{aligned}
\tmACK &= \stdname{aUnitBackoffPeriod}\\
       &\hspace{1em}+ \stdname{aTurnaroundTime}+\stdname{phySHRDuration}\\
       &\hspace{1em}+ \lceil 6 \cdot \stdname{phySymbolsPerOctet} \rceil\\
&= 20 + 12 + 10 + 6\cdot2 = 54.
\end{aligned}
\end{align}
This gives a total time for an unsuccessful transmission of
\begin{align}
\LenFail = \Len + \frac{\tmACK}{20} = \Len + 2.7.
\end{align}
time units. In relation to $\Sb$, the packet generation rate on each client in upstream direction is
\begin{align}
\pgenup = \frac{\Sb}{\Iup}
\end{align}
and since the gateway $\rootnode$ generates packets for each client its packet generation rate is
\begin{align}
\pgendown= \frac{\left(\Nodes-1\right) \cdot \Sb}{\Idown}.
\end{align}

\section{Traffic Distribution}
\label{sec:traffdist}
\begin{figure}[htb]
\centering
	\includegraphics[]{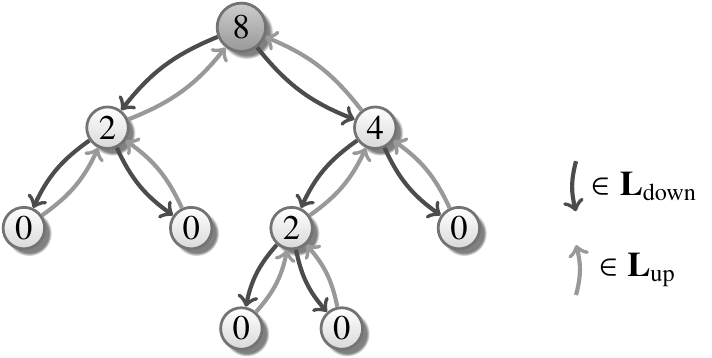}
\caption{An example for the set of active links and the values of \pdescn in the nodes.\label{fig:tree}}
\vspace{-0.4cm}
\end{figure}

The distribution of the traffic is directly calculated from the tree. $\linkrate{l}$ denotes the packet sending rate of the sender $v$ of a link $l = (v,w) \in \L$, including generated and forwarded traffic. At the receiver side $\forwardedl$ is the effective packet rate to be forwarded by $w$. Furthermore, \pdescn denotes the number of proper descendants of a nodes, that is the number of nodes whose traffic is routed through this node (Fig. \ref{fig:tree}).

\begin{figure}[htb]
\centering
\subfloat[Upstream]{\centering\includegraphics[]{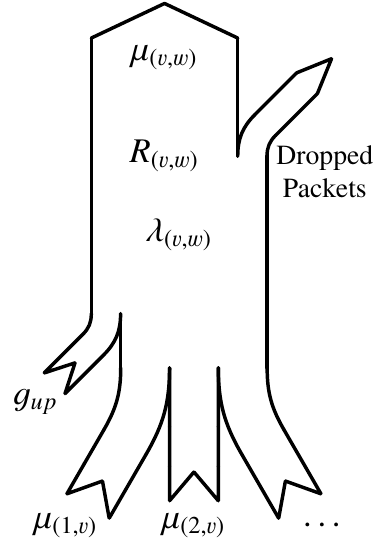}}\hspace{0cm}
\subfloat[Downstream]{\centering\includegraphics[]{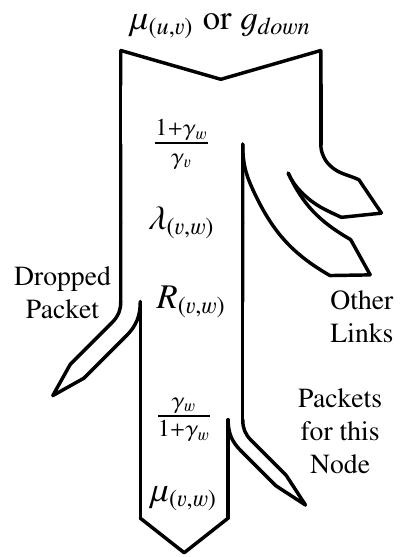}}
\caption{The data flow of a link.\label{fig:sankey}}
\vspace{-0.5cm}
\end{figure}

Fig. \ref{fig:sankey} depicts the possible inflows and drains of traffic. For upstream, $\linkrate{(v,w)}$ is the sum of the traffic of all proper descendants and the traffic generated in this node. Some packets will be dropped with a probability of $\R{(v,w)}$, because it was not possible to transmit them successfully.

For downstream, the gateway is the only node that autonomously generates packets with rate $\pgendown$. The respective receivers $v$ of a link $(u,v)$ transmit traffic to be forwarded to their proper descendants with rate $\forwarded{(u,v)}$. This traffic is split up between the outgoing links. Each link $(v,w)$ going out of $v$ gets that fraction of the traffic $\linkrate{(v,w)}$ that corresponds to the number of nodes reachable via this link $\left(1+\pdesc{w}\right)$ in relation to the total number of nodes reachable via $v$, namely $\pdesc{v}$. The second difference to upstream is the fact that each node will consume the fraction of the traffic intended for this node. Altogether, this results in the recursive expressions

\begin{align}
\linkrate{(v,w)} &= \begin{cases}
\frac{1+\pdesc{w}}{\pdesc{v}} \cdot \pgendown & \makebox[2cm][r]{$v = \rootnode \wedge (v,w) \in \Ldown$}\\
    \frac{1+\pdesc{w}}{\pdesc{v}} \cdot \forwarded{(u,v)} & \makebox[2cm][r]{$\T{u,v} \wedge (v,w) \in \Ldown$}\\
\pgenup + \sum_{(u,v)\,\in\,\L_{\mbox{\tiny up}}} \forwarded{(u,v)} & \makebox[2cm][r]{$(v,w) \in \Lup$}\end{cases},\\[1em]
\forwarded{(v,w)} &= \linkrate{(v,w)}\cdot \R{(v,w)}\cdot\begin{cases}\frac{\pdesc{w}}{1+\pdesc{w}} & \makebox[2.4cm][r]{$(v,w) \in \Ldown$}\\
1 & \makebox[2.4cm][r]{$(v,w) \in \Lup$}\end{cases}.
\end{align}

Since the probability of a pending packet is modeled as Poisson distributed, it is given by $\Psendl = 1 - e^{-\linkratel}$. 

\section{Model for IEEE 802.15.4 CSMA/CA}
\label{sec:csma}
\newcommand{\bnd}[1]{\nonumber&\makebox[\columnwidth][r]{$#1$}}%

The IEEE 802.15.4 wireless standard uses CSMA/CA for media access control. There are other options, such as a beacon enabled mode and guaranteed time slots (GTS), but they are not easily realizable in multi-hop networks.

The MAC layer is modeled by a Markov chain that is depicted in \fref{csma} with the following states.
\begin{itemize}[leftmargin=1.5cm,itemsep=0pt]
\item[$\mstate{idle}$] No transmission pending.
\item[$\mstate{i,k,j}$] In the $i$\textsuperscript{th} backoff stage after $j$ transmission attempts. CCA in $k$ steps.
\item[$\mstate{-1,q,j}$] During a successful transmission after $q$ time steps and $j$ preceding transmission attempts.
\item[$\mstate{-2,q,j}$] During a colliding transmission after $q$ time steps and $j$ preceding transmission attempts.
\end{itemize}
$\mtr{\xi}{\zeta}$ is the probability of going from $\xi$ to $\zeta$ in one step. The probability that the sender will leave the idle state within one time unit is given by $\Psendl$ as calculated in the previous section. Therefore, the probability of staying in the idle state is
\begin{align}
&\mmtr{idle}{idle} = 1-\Psendl. & \label{math:nonewdata}
\end{align}
Otherwise, the MAC layer starts the backoff period. That is it waits for an integer random time span. $\maxbackoffs$ is the constant parameter \stdname{macMaxCSMABackoffs} that corresponds to the maximum number of backoffs as defined in \cite{802154}. Note that, according to the standard, $\maxbackoffs$ does not include the first unconditional backoff, so the actual number of backoffs is $\maxbackoffs+1$. The maximal backoff time span $\Wi-1$ increases with the number $i \in [0,\maxbackoffs]$ of failed channel accesses, with 
\begin{align}
\W{i} = \begin{cases}
2^{\initbexp} & i = 0 \\ 
2^i\cdot \W{0} & 0 < i \leq \mo = \maxbexp-\initbexp\\
2^{\mo}\cdot \W{0} & i > \mo
\end{cases},
\end{align}
with the constant parameters \initbexp and \maxbexp that are the initial and maximum backoff exponents.
The die is symbolic for the associated choice. It is not a state in itself, but just bundles the different ways to go into the backoff period. The probability of choosing a specific state is $\Wi$, therefore
\begin{align}
    &\mmtr{idle}{0,k,0} = \frac{\Psendl}{\W{0}}, \label{math:newdata}&0 \leq k \leq \W{0}-1.
\end{align}
Within one time unit, the backoff duration is decreased by one
\begin{align}
&\mmtr{i,k+1,j}{i,k,j} = 1,  \label{math:backoff}\\
\bnd{0 \leq i \leq \maxbackoffs \wedge 0 \leq k < \W{i}-1 \wedge 0 \leq j \leq \maxretrans.}
\end{align}

\begin{figure}[htb]
\centering
\vspace{-0.5cm}
\includegraphics{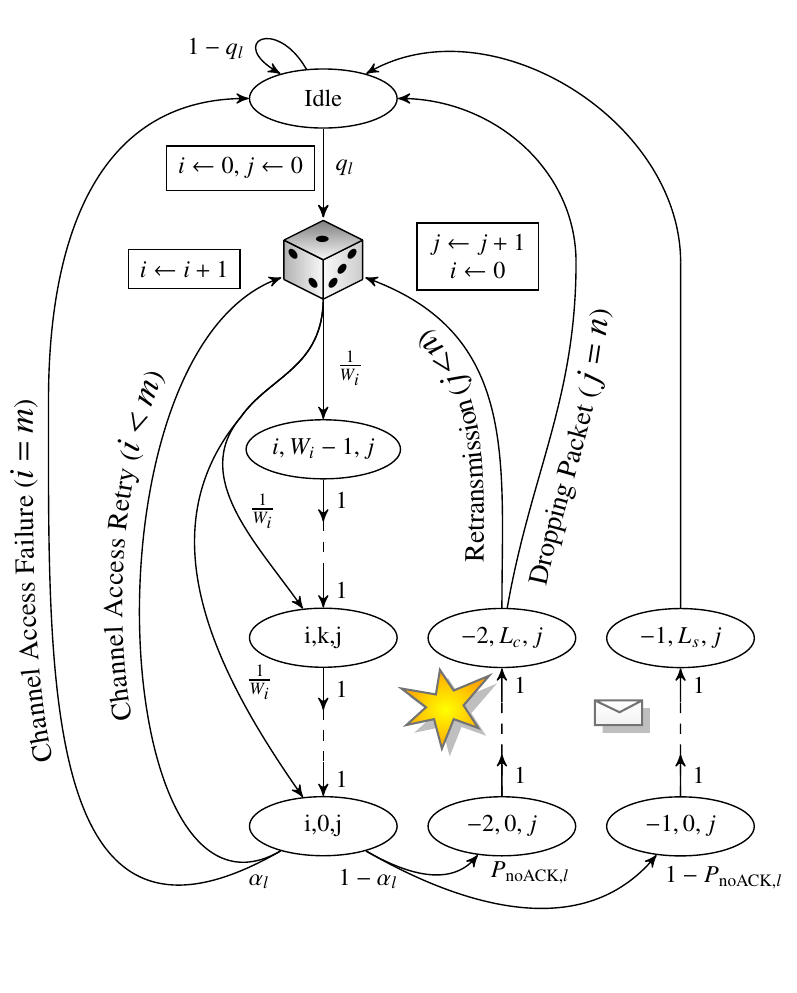}
\vspace{-0.5cm}
\caption{Sketch of the IEEE 802.15.4 CSMA/CA technique.\label{fig:csma}}
\vspace{-0.4cm}
\end{figure}%

As soon as the state $\mstate{i,0,j}$ is reached, channel sensing takes place. If the channel is sensed busy with probability $1-\al$ as calculated in Sect. \ref{sec:neighbour}, the channel access retry counter $i$ is incremented and the backoff procedure is started again
\begin{align}
    &\mmtr{i-1,0,j}{i,k,j} = \frac{\al}{\W{i}}, \label{math:newcca}\\
    \bnd{0 < i \leq \maxbackoffs \wedge 0 \leq k \leq \W{0}-1 \wedge 0 \leq j \leq \maxretrans.}
\end{align}
If $i$ has reached $\maxbackoffs$, the packet will be dropped
\begin{align}
&\mmtr{\maxbackoffs,0,j}{idle} = \al, \label{math:idlecf} & 0 \leq j \leq \maxretrans.
\end{align}

Otherwise, if the channel is sensed free, the transmission takes place. If the packet or the associated acknowledgement sent out by the receiver will eventually collide, $\mstate{-2,0,j}$ is chosen as next state
\begin{align}
&\mmtr{i,0,j}{-2,0,j} = (1-\al)\cdot \Pnoack, \label{math:collide}\\
\bnd{0 \leq i \leq \maxbackoffs \wedge 0 \leq j \leq \maxretrans.}
\end{align}
Of course, the real hardware does not know if the transmission will collide, but in the analytical model the probability $\Pnoack$ can be computed from the collision graph as presented in the next section. In case of a upcoming successful transmission, the $\mstate{-1,0,j}$ state is chosen
\begin{align}
&\mmtr{i,0,j}{-1,0,j} = (1-\al) \cdot\left(1 - \Pnoack\right), \label{math:notcollide}\\
\bnd{0 \leq i \leq \maxbackoffs \wedge 0 \leq j \leq \maxretrans.}
\end{align}
During the transmission, the counter $h$ is increased until the time is over
\begin{align}
\mmtr{-1,h,j}{-1,h+1,j} = 1,\,&0 \leq h < \LenFail \wedge 0 \leq j \leq \maxretrans,\label{math:sending}\\
\mmtr{-2,h,j}{-2,h+1,j} = 1,\,&0 \leq h < \LenSuccess \wedge 0 \leq j \leq \maxretrans.\label{math:colliding}
\end{align}
After a collision, the retransmission counter $j \in [0,\maxretrans]$, with $\maxretrans$ being the maximum number of retries $\stdname{macMaxFrameRetries}$, is incremented, the channel access retry counter $i$ is reset to zero and a new backoff period starts
\begin{align}
    &\mmtr{-2,L_c,j-1}{0,k,j} = \frac{1}{\W{0}}, &0 < j \leq \maxretrans.\label{math:retry}
\end{align}
Finally, the idle state is reached after a successful transmission or as soon as the maximum number of retransmissions is reached and the packet is dropped
\begin{align}
&\mmtr{-1,L_s,j}{idle} = 1,\label{math:success} & 0 \leq j \leq \maxretrans,\\
&\mmtr{-2,L_c,n}{idle} = 1.\label{math:idlecr}
\end{align}
All transition probabilites not specified before are zero. The stationary distribution of the Markov chain yields the probability that a given link is in a specific state as derived in Appendix~\ref{sec:stationary}. In particular, the probability of being in a generic sensing state is given by
\begin{align}
\tl &= \stat{0,0,0} \geo{\al}{\maxbackoffs+1} \geo{\yl}{n+1},\label{math:tl}
\end{align}
with the shortcut $\yl \coloneqq \Pnoack \cdot \left(1-\al^{\maxbackoffs+1}\right)$ and $\stat{0,0,0}$, the probability of being in the first channel sensing state, given by %
\begin{align}
&\frac{1}{\stat{0,0,0}} = \frac{1}{2} \left( W_0 \geok{2\cdot\al}{\min(\maxbackoffs,\mo)+1} + \geo{\al}{\min(\maxbackoffs,\mo)+1}\right.\nonumber\\
&\makebox[\columnwidth][r]{$\displaystyle\left. + \frac{\left(2^{\maxbexp} + 1\right) \al^{\mo+1} \left(1 -  \al^{\max(0,\maxbackoffs - \mo)} \right)}{1-\al} \right)\geo{\yl}{n+1}$}\nonumber\\
&+ (1-\al^{\maxbackoffs+1}) \geo{\yl}{n+1} \left( L_s\cdot\left(1-\Pnoack\right) + L_c\cdot\Pnoack\right)\nonumber\\
&+ \frac{1}{\Psendl} \left( \yl^{n+1}\tight+\tight \geo{\yl}{n+1} \left(\al^{\maxbackoffs+1}\tight+\tight\frac{1 - \Pnoack}{\left(1-\al^{\maxbackoffs+1}\right)^{-1}}\right)\right).\label{math:b000}
\end{align}

\section{Simultaneous Transmissions}
\label{sec:coll}
The probability that the sender of a link $j \in L$ does not start a transmission at a given point in time is given by
\begin{align}
\t{j}\a{j}+\left(1-\t{j}\right).
\end{align}
with $\t{j}$, the probability that the sending node of $j$ is trying to access the channel and $\a{j}$, the probability that the channel would be sensed busy (see Sect. \ref{sec:csma}).
So either the sender tries to access the channel, but senses it busy $\left(\t{j}\a{j}\right)$ or it does not try to access the channel at all $\left(1-\t{j}\right)$. For a given set of links $\sublinks \subset \L$, the probability that at least one $j \in \sublinks$ starts a transmission is therefore given by the complementary event of the event that all nodes are not sending
\begin{align}
\somesending{\sublinks} &= 1 - \prod_{j\,\in\,\sublinks} \left( \t{j}\a{j}+\left(1-\t{j}\right) \right).
\end{align}

The same expression can be derived from the expression $\mathrm{Pr}\left[\mathcal{A}_l\right]$ in~\cite{dimarco_analytical_2012} as conducted in Appendix \ref{sec:alternativeSomesending}, but is less complex and takes much less computation time. For time intervals of $t \geq 1$, we take into account that the random variables are Poisson distributed, so
\begin{align}
	\somesending{\sublinks} = 1 - e^{-\sigma}\quad \Leftrightarrow \quad e^{-\sigma} = 1 - \somesending{\sublinks}
\end{align}
for some rate $\sigma$. For larger intervals, the rate increases proportional, so the probability of at least one transmission start within this time interval is
\begin{align}
	\someoccupy{t}{\sublinks} &= 1 - e^{-\sigma\cdot t}\\
	& = 1 - \left(1 - \somesending{\sublinks}\right)^{t}.
\end{align}

\section{Analyzing the Neighborhood}
\label{sec:neighbour}
The formula derived in the previous section can be used to calculate collisions that occur if at least two transmissions arrive at the same receiver at the same time. In order to represent packets (sent along a link) as well as acknowledgements (sent in the reverse direction), even under the hidden node problem, an enhanced conflict graph is introduced that considers all constallations of senders (in the following: sends a packet, but receives an acknowledgement) and receivers (receives a packet, but sends an acknowledgement).

\begin{figure}[htb]
\vspace{-0.2cm}
\centering
	\includegraphics[]{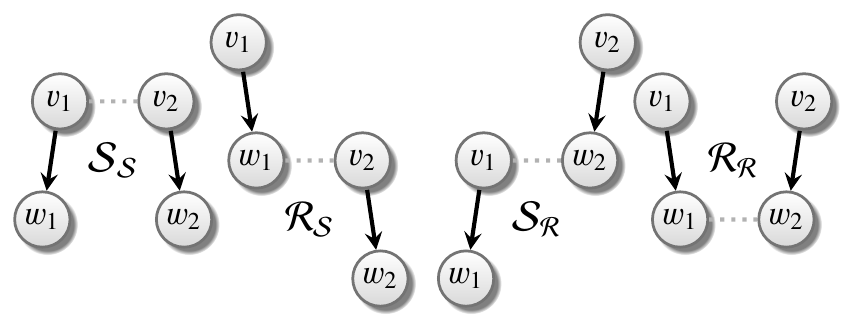}
\vspace{-0.1cm}
\caption{The four basic relations for conflicts of two links.\label{fig:rel}}
\vspace{-0.2cm}
\end{figure} %

For a link $(v_1,w_1) \in \L$ the possibly disturbing links $(v_2,w_2) \in \L$ are expressed in terms of the basic constellations as depicted in \fref{rel} and formulated as %
\begin{align}
    \SS{(v_1,w_1)} &= \linkset{(v_2,w_2) \in \L}{\inrange{v_1,v_2} \wedge v_1 \neq v_2},
\intertext{that is the sender of one link receives transmissions from the sender of another link,}
\RS{(v_1,w_1)} &= \linkset{(v_2,w_2) \in \L}{\inrange{w_1,v_2} \wedge v_1 \neq v_2},
\intertext{that is the receiver of one link receives transmissions from the sender of another link,}
\SR{(v_1,w_1)} &= \linkset{(v_2,w_2) \in \L}{\inrange{v_1,w_2} \wedge v_1 \neq v_2},
\intertext{that is the sender of one link receives transmissions from the receiver of another link,}
\RR{(v_1,w_1)} &= \linkset{(v_2,w_2) \in \L}{\inrange{w_1,w_2} \wedge v_1 \neq v_2},
\end{align}%
that is the receiver of one link receives transmissions from the receiver of another link.

In the following, $\collpacketl$ denotes the event that a packet collides on a link $l$, while $\collackl$ denotes the event that an acknowledgement collides. The constellation that the sender $v_2$ of the link $j = \left(v_2,w_2\right)$ is in the range of both the sender $v_1$ and the receiver $w_1$ of a link $l = \left(v_1,w_1\right)$ can be expressed as
\begin{align}
j \in \RS{l} \cap \SS{l}.
\end{align}
In this constellation, both senders will wait for each other if they sense an ongoing transmission. Although, a sender might sense within the turnaround time of the other one and vice versa and they therefore start transmitting within an interval of 2 time units. So the probability of a collision of two packets on links that obey this constellation is expressed as
\begin{align}
\pnsp{0} &= \someoccupy{2}{\RS{l} \cap \SS{l}}.\label{math:fullcoverage}
\end{align}
If receiver $w_1$ might receive a packet from sender $v_2$, but the transmission of $v_2$ is not recognized by $v_1$, the channel sensing has basically no effect. This is called the hidden node problem. Two transmissions of length $\Len$ might overlap in an interval of $2\Len$. This is calculated as
\begin{align}
\pnsp{1} &= \someoccupy{2\Len}{\RS{l} \setminus \SS{l}}.
\end{align}
The transmission can not only collide with another packet, but also with an acknowledgment. If all involved nodes are in range, the new packet might collide with an acknowledgement if the carrier sensing takes place between the arrival of the original packet and the beginning of the acknowledgment
\begin{align}
\pnsp{2} &= \someoccupy{1}{\SS{l} \cap \SR{l} \cap \RR{l}}.\label{math:pns3}
\end{align}
If the senders can not hear each other, the sensing might also take place at the end of the packet,
\begin{align}%
\pnsp{3} &= \someoccupy{2}{\left(\SR{l} \cap \RR{l}\right) \setminus \SS{l}}.
\end{align}
If it were to occur earlier, the transmission would collide with the packet, so it should not be counted here, too. If the acknowledgement can not be heared, the packet transmission might take place during the whole acknowledgement
\begin{align}%
\pnsp{4} &= \someoccupy{\Lack}{\left(\SS{l} \cap \RR{l}\right) \setminus \SR{l}}.
\end{align}
This case is extended in the following constellation by the sensing before the pause $\left(\mbox{cf. }\pnsp{3}\right)$
\begin{align}%
\pnsp{5} &= \someoccupy{\Lack+1}{\left(\RS{l} \cap \RR{l}\right) \setminus \SS{l} \setminus \SR{l}}.
\end{align}
Finally, the packet might overlap with an acknowledgement during the whole transmission, if the sensing of both senders is not effective
\begin{align}%
    \pnsp{6} &= \someoccupy{\Len+\Lack}{\RR{l} \setminus \SS{l} \setminus \SR{l} \setminus \RS{l}}.
\end{align}
In addition, the acknowledgement itself might be affected by collision with a packet
\begin{align}
\pnsack{0} &= \someoccupy{1}{\SS{l} \cap \RS{l}},
\end{align}
and in particular if the acknowledgement can not be heared
\begin{align}
\pnsack{1} &= \someoccupy{\Lack}{\SS{l} \setminus \RS{l}}.
\end{align}

The probability that a packet collides with at least one other packet or acknowledgement is
\begin{align}
\Prob{\collpacketl} &= P\left(\bigcup_{i=0}^6 \collpacket{l,i}\right).
\end{align}
Note that the events are not mutually exclusive, so inclusion-exclusion principle has to be applied for the calculation. Furthermore, a transmission of $b$ bytes might be dropped because of a bad link with probability $\PER{b,l}$ as given in (\ref{eq:per}). Taking this into account, the overall probability that a packet is not successfully received is
\begin{align}
	\Prob{\lostpacketl} &= \Prob{\collpacketl} + (1-\Prob{\collpacketl}) \cdot \PER{\BytesPacket,l}.
\end{align}
The combined probability of a lost acknowledgement is
\begin{align}
	\Prob{\collackl} = P\left(\bigcup_{i=0}^{1} \collack{l,i}\right).
\end{align}
Taking a possible bad link into account, the probability of losing the acknowledgment is
\begin{align}
    \Prob{\lostackl} &= \Prob{\collackl} + \left(1-\Prob{\collackl}\right) \cdot \PER{\BytesACK,l}.
\intertext{Finally, the overall probability that no acknowledgement arrives at the sender is}
\Pnoack &= \Prob{\lostpacketl} + \left(1-\Prob{\lostpacketl}\right)\cdot\Prob{\lostackl}.
\end{align}%

The probability $\al$ that the channel is sensed busy is calculated along the same line as the probability that either a packet transmission or an acknowledgment arrives at the sender
\begin{align}
	\apkt &= \someoccupy{\Len}{\SS{l}},\\
	\aack &= \someoccupy{\Lack}{\SR{l}},\\
	\al &= \apkt + \aack - \apkt \cdot \aack.
\end{align}

Note that all sets can be computed offline, so the actual computation is linear with the number of neighbours.

\section{Link Reliability Considering Retransmissions}
It turned out that the reliability of packet transmission is not independent of the current retransmission attempt and this effect has a major impact on the results \cite{meiermodel}. Therefore, the model contains an elaborate handling of this matter.

The probability of a mutual disturbance of at least two hidden nodes is calculated 
\begin{align}
\pnsb{2} = \someoccupy{2 \Len+2}{\left( \RS{l} \cap \SR{l} \right) \setminus \SS{l}}.
\end{align}
After such a mutual disturbance took place, those nodes will issue a retransmission. The backoff exponent is reset to the initial backoff counter $\initbexp$, so there are $W_0 = 2^{\initbexp}$ possible backoff time spans. The event that the retransmission collides again, given $i$ other nodes with pending retransmissions is denoted as $\bc{i}$. For two nodes the probability is%
\begin{align}
\begin{aligned}
    \omega &= \min\left(\W{0} - \Len - 1,0\right)\\
	P\left(\bc{1}\right) &= 1 - \sum_{h=1}^{\omega} \frac{2h}{\W{0}^2} = 1 - \frac{\omega + \omega^2}{\W{0}^2}.
\end{aligned}
\end{align}
Of course, this expression is only a lower bound of the actual probability that there will be a repeated collision. For example, there might be several initial collisions at the same time leading to a higher probability of a repeated collision. It can be calculated by summing over all possible combinations of mutual disturbance. However, the price for this gain of accuracy is too high considering the exponential growth in computational complexity, in particular for large scale networks. Analogous to \mref{fullcoverage}, even senders which can mutually sense their transmissions might be affected by packet collision. The probability that at least two senders which can mutually sense their transmissions are affected by mutual disturbance is
\begin{align}
\pnsb{1} = \someoccupy{2}{\RS{l} \cap \SR{l} \cap \SS{l}}.
\end{align}
$\bsc{j}$ is defined analogous to $\bc{i}$ and the corresponding probability of a repeated collision of two nodes is calculated as
\begin{align}
    P\left(\bsc{1}\right) = \W{0}^{-1}.
\end{align}

With these quantities, an absorbing Markov chain is built with the states
\begin{itemize}[leftmargin=1.5cm,itemsep=0pt,topsep=-1pt]
\item[$\rstsucc$] Successful transmission
\item[$\rstcf$] Channel access failure
\item[$\rstate{0,0}$] No preceding mutual disturbance
\item[$\rstate{1,0}$] Hidden node(s) with pending retransmission
\item[$\rstate{0,1}$] Visible node(s) with pending retransmission
\item[$\rstate{1,1}$] Hidden node(s) and visible node(s) with pending retransmission
\end{itemize}
In the following, all unspecified transition probabilities as well as $P\left(\bc{0}\right)$ and $P\left(\bsc{0}\right)$ are zero.
$\rstsucc$ and $\rstcf$ are the absorbing final states, so
\begin{align}
\mtr{\rstsucc}{\rstsucc} &= 1,\\
    \mtr{\rstcf}{\rstcf} &= 1.
\end{align}
The probability of a channel access failure is always the same
\begin{align}
\mtr{\rstate{p,q}}{\rstcf} &= \alpha_l^{\maxbackoffs+1}.
\end{align}
All following transitions need to take this into account with
\begin{align}
\bl = 1 - \alpha_l^{\maxbackoffs+1}.
\end{align}
The probability of a successful transmission is given by the probability that neither a conventional collision takes place with $\Prob{\lostpacketl}$ nor a repeated collision
\begin{align}
    \mtr{\rstate{p,q}}{\rstsucc} &= \bl \cdot \left(1 - P\left(\lostpacketl \,\cup\, \bc{p} \,\cup\, \bsc{q} \right)\right).
\end{align}
Note that $\Prob{\lostpacketl}$ is used instead of $\Pnoack$, because even if the acknowledgement does not arrive at the sender, the packet itself might be transmitted successfully. The transmission to $\rstate{0,0}$ takes place if no repeated collision takes place, but the packet collides anyway with another transmission
\begin{align}
	\mtr{\rstate{p,q}}{\rstate{0,0}} &= \bl \cdot \left( \Prob{\lostpacketl} - P\left(\bothset \cup \bothsameset\right) \right) \\\nonumber
                                 &\quad\quad\cdot\left(1 - P\left(\bc{p} \cup \bsc{q}\right)\right).
\end{align}
All remaining state transitions indicate that any repeated collision takes place
\begin{align}
\mtr{\rstate{p,q}}{\rstate{1,0}} &= \bl \cdot
    P\left(\bothset \,\cup\, \bc{p}\right)\\\nonumber
&\hspace{1em}\cdot
\left(1-P\left(\bothsameset \,\cup\, \bsc{q}\right)\right),\\
\mtr{\rstate{p,q}}{\rstate{0,1}} &= \bl \cdot
\left(1 - P\left(\bothset \,\cup\, \bc{p}\right)\right)\\\nonumber
&\hspace{1em}\cdot
P\left(\bothsameset \,\cup\, \bsc{q}\right),\\
\mtr{\rstate{p,q}}{\rstate{1,1}} &= \bl \cdot
P\left(\bothset \,\cup\, \bc{p}\right)\\\nonumber
&\hspace{1em}\cdot
P\left(\bothsameset \,\cup\, \bsc{q}\right).
\end{align}

The probability $\Rl$ that a packet was successfully transmitted is described by the probability of reaching \rstsucc from $\rstate{0,0}$ in at most $\maxretrans+1$ steps, calculated by the corresponding power of the transition matrix.

\section{Path Reliability}
The end-to-end reliability in upstream direction $\Rup{n}$ is the probability that a packet sent by a node $n$ arrives at the gateway. It is given by the product of the reliabilities along the routing path
\begin{align}
\Rup{c} = \begin{cases}
    \Rup{p} \cdot \R{(c,p)} & c \neq \rootnode \wedge (c,p) \in \Lup\\ 
1 & c = \rootnode\\
\end{cases}.\label{math:Rupc}
\end{align}
By construction, there is always exactly one $p$ with $(c,p) \in \Lup$ unless $c$ is the gateway. The reliability in downstream direction $\Rdown{n}$, that is the probability that a packet sent by the gateway arrives at its destination, is given correspondingly by multiplication of the reliabilities in downstream direction
\begin{align}
    \Rdown{c} = \begin{cases}
        \Rdown{p}\cdot \R{(p,c)} & c \neq \rootnode \wedge (p,c) \in \Ldown\\
1 & c = \rootnode\\
\end{cases}.
\end{align}

\addcontentsline{toc}{section}{List of Symbols}
\printglossary[type=symbols]
%
%
\IEEEtriggeratref{7}
\bibliographystyle{IEEEtran}
	
\bibliography{model}

\appendices
\onecolumn

\section{Stationary Distribution}
\label{sec:stationary}
In the asymptotic case, the probability $\stat{\zeta}$ of being in state $\mstate{\zeta}$ is calculated by summing over the probabilities of the preceding states $\Xi$, weighted by the transition probability.%
\begin{align}
    \stat{\zeta} &= \sum_{\xi \in \Xi} \stat{\xi} \cdot \mtr{\mstate{\xi}}{\mstate{\zeta}}
\end{align}

Therefore, from \mref{newdata} and \mref{backoff} results
\begin{align}
    \stat{0,k,0} &= \frac{\ql}{\W{0}} \stat{idle} + \stat{0,k+1,0}, &0 \leq k < \W{i} - 1.
    \intertext{Since the maximum backoff period is $\W{i}-1$, it holds}
    \stat{i,\W{i},j} &= 0 & 0 \leq i \leq \maxbackoffs \wedge 0 \leq j \leq \maxretrans,\\
    \intertext{so the recursion resolves to}
    \stat{0,0,0} &= \W{0} \cdot \frac{\ql}{\W{0}} \stat{idle} = \ql \cdot \stat{idle}.\label{math:000rec}
    \intertext{The same for \mref{newcca} results in} 
    \stat{i,0,j} &= \W{i} \cdot \frac{\al}{\W{i}} \stat{i-1,0,j} = \al \cdot \stat{i-1,0,j}, & 0 < i < \maxbackoffs - 1 \wedge 0 \leq j < \maxretrans,\label{math:i0jrec}\\
                 &= \al^i \cdot \stat{0,0,j} \label{math:bi0j}
    \intertext{and for \mref{retry}}
    \stat{0,0,j} &= \W{0} \cdot \frac{1}{\W{0}}\stat{-2,\LenFail,j-1} = \stat{-2,\LenFail,j-1}, & 0 < j < \maxretrans.\label{math:00jrec}
\intertext{together with \mref{collide}}
\stat{0,0,j} &= \sum_{i=0}^{\maxbackoffs}\stat{i,0,j-1} \cdot (1-\al)\cdot \Pnoack \\
&= \stat{0,0,j-1} \cdot  (1-\al)\cdot \Pnoack \cdot \sum_{i=0}^{m} \al^i\\
&= \stat{0,0,0} \cdot \left((1-\al)\cdot \Pnoack \cdot \sum_{i=0}^{m} \al^i\right)^j\\
\intertext{and taking the sum as the partial sum of the geometric series}
&= \stat{0,0,0} \cdot \left((1-\al)\cdot \Pnoack \cdot \geo{\al}{\maxbackoffs+1}\right)^j\\
&= \stat{0,0,0} \cdot \left(\Pnoack \cdot \left(1-\al^{m+1}\right)\right)^j\\
&= \stat{0,0,0} \cdot \yl^j. \label{math:b00j}
\end{align}

For each $k \in [0,W_{i} - 1] \wedge i \geq 0$ holds with \mref{newdata}, \mref{backoff}, \mref{newcca} and \mref{retry}
\begin{align}
\stat{i,k,j} &= \begin{aligned}\stat{i,k+1,j} + \begin{cases}
		\frac{\ql}{W_0} \stat{idle} & \mbox{for } i = 0 \wedge j = 0\\
		\frac{\al}{W_0} \stat{i-1,0,j} & \mbox{for } 0 < i < m-1 \wedge 0 \leq j < n\\
		\frac{1}{W_0} \stat{-2,L_c,j-1} & \mbox{for } i = 0 \wedge 0 < j < n
		\end{cases} \label{math:backoffsum}
\end{aligned}
\intertext{and taking \mref{000rec}, \mref{i0jrec} and \mref{00jrec} into account}
&= \begin{aligned}\stat{i,k+1,j} + \begin{cases}
		\frac{\ql}{W_0} \frac{1}{\ql} \stat{0,0,0} & \mbox{for } i = 0 \wedge j = 0\\
		\frac{\al}{W_i} \frac{1}{\al} \stat{i,0,j} & \mbox{for } 0 < i < m-1 \wedge 0 \leq j < n\\
		\frac{1}{W_0} \stat{0,0,j} & \mbox{for } i = 0 \wedge 0 < j < n
		\end{cases}
\end{aligned}\\
&= \begin{aligned}\stat{i,k+1,j} + \stat{i,0,j}\cdot\frac{1}{W_{i}}.
\end{aligned}\\
\intertext{The recursion dissolves to}
\stat{i,k,j} &= \stat{i,0,j}\cdot\frac{W_{i} - k}{W_{i}},\\
\intertext{and together with \mref{bi0j} and \mref{b00j}}
\stat{i,k,j} &= \stat{0,0,0} \cdot \yl^j \cdot \al^i \cdot\frac{W_{i} - k}{W_{i}}.\label{math:bikj}
\end{align}

The probabilities in the stationary distribution have to sum up to one (normalization condition), that is
\begin{align}
   \sum_{i=0}^{\maxbackoffs} \sum_{j=0}^n \sum_{k=0}^{W_{i}-1} \stat{i,k,j} + \sum_{j=0}^n \left(\sum_{h=0}^{L_s-1} \stat{-1,h,j} + \sum_{h=0}^{L_c-1} \stat{-2,h,j}\right) + \stat{idle}  = 1.\label{math:normalization}
\end{align}
From \mref{bikj} results
\begin{align}
    \sum_{i=0}^{\maxbackoffs} \sum_{j=0}^n \sum_{k=0}^{W_{i}-1}\stat{i,k,j} &= \sum_{i=0}^{\maxbackoffs} \sum_{j=0}^n \sum_{k=0}^{\W{i}-1}\stat{0,0,0} \cdot \yl^j \cdot \al^i \cdot\frac{W_{i} - k}{W_{i}}\\
&= \stat{0,0,0} \sum_{i=0}^{\maxbackoffs} \sum_{j=0}^n \yl^j \cdot \al^i \cdot \sum_{k=0}^{W_{i}-1} \frac{W_{i} - k}{W_{i}}\\
&= \stat{0,0,0} \sum_{j=0}^n \yl^j \cdot \sum_{i=0}^{\maxbackoffs} \al^i \cdot \frac{W_{i} + 1}{2}\\
    \intertext{splitting the second sum at \mo}
    &= \stat{0,0,0}  \sum_{j=0}^n \yl^j \left(\sum_{i=0}^{\min(\maxbackoffs,\mo)} \al^i \cdot \frac{\W{i} + 1}{2} + \sum_{i=\mo+1}^{\maxbackoffs} \al^i \cdot \frac{\W{i} + 1}{2}\right) \\
    &= \stat{0,0,0}  \sum_{j=0}^n \yl^j \left(\sum_{i=0}^{\min(\maxbackoffs,\mo)} \al^i \cdot \frac{2^i \W{0} + 1}{2} + \sum_{i=\mo+1}^{\maxbackoffs} \al^i \cdot \frac{2^{\mo} \W{0} + 1}{2}\right) \\
    &= \frac{\stat{0,0,0}}{2} \geo{\yl}{n+1} \left( \sum_{i=0}^{\min(\maxbackoffs, \mo)} \al^i \cdot \left(2^i \W{0} + 1\right) + \sum_{i=\mo+1}^{\maxbackoffs} \al^i \cdot \left(2^{\mo} 2^{\initbexp} + 1\right) \right)\\
    &= \frac{\stat{0,0,0}}{2} \geo{\yl}{n+1} \left(\sum_{i=0}^{\min(\maxbackoffs, \mo)} \left(2\al\right)^i \cdot  W_0  + \sum_{i=0}^{\min(\maxbackoffs, \mo)} \al^i + \left(2^{\maxbexp-\initbexp+\initbexp} + 1\right) \sum_{i=\mo+1}^{\maxbackoffs} \al^i \right)\\
    &= \frac{\stat{0,0,0}}{2} \geo{\yl}{n+1} \left(\sum_{i=0}^{\min(\maxbackoffs, \mo)} \left(2\al\right)^i \cdot  W_0  + \sum_{i=0}^{\min(\maxbackoffs, \mo)} \al^i + \left(2^{\maxbexp} + 1\right) \sum_{i=\mo+1}^{\maxbackoffs} \al^{\mo+1}\al^{i-\mo-1} \right)\\
&= \frac{\stat{0,0,0}}{2} \geo{\yl}{n+1} \left( \W0 \geok{2\al}{\min(m,\mo)+1} + \geo{\al}{\min(m,\mo)+1}
+ \left(2^{\maxbexp} + 1\right) \al^{\mo+1} \sum_{i=0}^{m-\mo-1} \al^i\right)\\
&= \frac{\stat{0,0,0}}{2} \geo{\yl}{n+1} \left( \W0 \geok{2\al}{\min(m,\mo)+1} + \geo{\al}{\min(m,\mo+1)} \right.\nonumber\\
&\left.\hspace{23.3em} + \frac{\left(2^{\maxbexp} + 1\right) \al^{\mo+1} \left(1 -  \al^{\max(0,m - \mo)} \right)}{1-\al} \right).\label{math:triplesum}
\end{align}

It follows from \mref{notcollide} and \mref{sending}
\begin{align}
\stat{-1,h,j} &= \stat{-1,0,j} = \sum_{i=0}^{\maxbackoffs} \left(1-\al\right) \cdot \left(1-\Pnoack\right) \cdot \stat{i,0,j}\\
              &=  \left(1-\Pnoack\right) \cdot \left(1-\al\right) \cdot \stat{0,0,0}\cdot \yl^j \cdot \sum_{i=0}^{\maxbackoffs} \al^i\\
              &= \left(1-\Pnoack\right)\cdot \stat{0,0,0} \cdot \yl^j \cdot\left(1-\al^{\maxbackoffs+1}\right) \\
\intertext{and analog from \mref{collide} and \mref{colliding}}
\stat{-2,h,j} &= \Pnoack \cdot \stat{0,0,0} \cdot \yl^j \cdot \left(1-\al^{\maxbackoffs+1}\right)\\
              &= \yl^{j+1} \cdot \stat{0,0,0}.
\intertext{Inserting these into the second term of the normalization condition gives}
\sum_{j=0}^n \left(\sum_{h=0}^{L_s-1} \stat{-1,h,j} + \sum_{h=0}^{L_c-1} \stat{-2,h,j} \right) &=\sum_{j=0}^{\maxretrans}\left(\LenSuccess \cdot \left(\left(1-\Pnoack\right)\cdot \stat{0,0,0} \cdot \yl^j \cdot\left(1-\al^{\maxbackoffs+1}\right)  \right) \right.\nonumber\\
&\hspace{4.4em}\left.+ \LenFail \cdot \left(\Pnoack \cdot \stat{0,0,0} \cdot \yl^j \cdot \left(1-\al^{\maxbackoffs+1}\right) \right)
\right)\\
&=\stat{0,0,0} \cdot \left(1-\al^{\maxbackoffs+1}\right) \cdot \left( L_s\cdot\left(1-\Pnoack\right) + L_c\cdot\Pnoack\right) \cdot \sum_{j=0}^{n} \yl^j \\
&=\stat{0,0,0} \cdot \left(1-\al^{\maxbackoffs+1}\right) \cdot \left( L_s\cdot\left(1-\Pnoack\right) + L_c\cdot\Pnoack\right) \cdot \geo{\yl}{n+1}.\label{math:sendsums}
\end{align}
\begin{align}
\intertext{The idle state probability is according to \mref{nonewdata}, \mref{idlecf}, \mref{success} and \mref{idlecr} given as}
\stat{idle} &= \left(1-\Psendl\right)\cdot\stat{idle}+\sum_{j=0}^n \left(\al\cdot \stat{m,0,j}+\stat{-1,L_s,j}\right)+\stat{-2,L_c,n}\\
    \stat{idle} - \left(1-\Psendl\right)\cdot\stat{idle} &= \sum_{j=0}^n \left(\al\cdot \al^{\maxbackoffs}\cdot \yl^j \cdot \stat{0,0,0}+\left(1-\Pnoack\right)\cdot \stat{0,0,0} \cdot \yl^j \cdot\left(1-\al^{\maxbackoffs+1}\right)\right)+\yl^{n+1}\cdot \stat{0,0,0}\\
    \Psendl\cdot\stat{idle} &= \stat{0,0,0}\cdot\left(\sum_{j=0}^n  \yl^j\left(\al^{\maxbackoffs+1} +\left(1-\Pnoack\right)\left(1-\al^{\maxbackoffs+1}\right)\right)+\yl^{n+1}\right)\\
\stat{idle} &= \frac{\stat{0,0,0}}{\Psendl}\cdot\left(\geo{\yl}{\maxretrans+1}\left(\al^{\maxbackoffs+1} + \left(1 - \Pnoack\right)\left(1-\al^{\maxbackoffs+1}\right)\right)+\yl^{n+1}\right).\label{math:idle}
\end{align}%

Finally, inserting \mref{triplesum}, \mref{sendsums} and \mref{idle} into the normalization condition \mref{normalization} gives
\begin{align}
    \stat{0,0,0}^{-1} &=  
    \frac{1}{2} \geo{\yl}{n+1} \left( \W0 \geok{2\al}{\min(m,\mo)+1} + \geo{\al}{\min(m,\mo+1)} + \frac{\left(2^{\maxbexp} + 1\right) \al^{\mo+1} \left(1 -  \al^{\max(0,m - \mo)} \right)}{1-\al} \right)\nonumber\\
&\hspace{1em}+\left(1-\al^{\maxbackoffs+1}\right) \cdot \left( L_s\cdot\left(1-\Pnoack\right) + L_c\cdot\Pnoack\right) \cdot \geo{\yl}{n+1}\nonumber\\
&\hspace{1em} + \frac{1}{\Psendl}\cdot\left(\yl^{n+1} + \geo{\yl}{\maxretrans+1}\left(\al^{\maxbackoffs+1} + \left(1 - \Pnoack\right)\left(1-\al^{\maxbackoffs+1}\right)\right)\right).
\end{align}

The probability $\tl$ is then calculated by summing over all $\stat{i,0,j}$ with $0 \leq i \leq \maxbackoffs$ and $0 \leq i \leq n$, giving
\begin{align}
\tl &= \sum_{i=0}^{m} \sum_{j=0}^{n} \stat{i,0,j}\\
&= \sum_{i=0}^{m} \sum_{j=0}^{n} \al^i \stat{0,0,j}\\
&= \stat{0,0,0} \sum_{i=0}^{m} \al^i \sum_{j=0}^{n} \yl^j \\
&= \stat{0,0,0} \geo{\al}{m+1} \geo{\yl}{n+1}.
\end{align}

\clearpage
\section{Alternative Derivation of $\somesending{\sublinks}$}
\label{sec:alternativeSomesending}
The expression in~\cite{dimarco_analytical_2012} iterates over all combinations of one up to all neighbouring links $l \in \sublinks$. This can also described as the power set of $\sublinks$ excluding the empty set 
\begin{align}
\mathcal{P}\left(\sublinks\right) \setminus \emptyset &= \left\{M | M \subseteq \sublinks \wedge M \neq \emptyset \right\}.
\end{align}

With this, the expression can be written as
\begin{align}
\somesending{\sublinks} &= \sum_{M \in \mathcal{P}\left(\sublinks\right) \setminus \emptyset} \left(\prod_{i \in M} \tau_i\right) \left( \prod_{i \in \sublinks \setminus M} \left(1 - \tau_i\right) \right) \left(1 - \prod_{i \in M} \alpha_i\right)\\
&= \sum_{M \in \mathcal{P}\left(\sublinks\right)} \left(\prod_{i \in M} \tau_i\right) \left( \prod_{i \in \sublinks \setminus M} \left(1 - \tau_i\right) \right) \left(1 - \prod_{i \in M} \alpha_i\right)
- \left(\prod_{i \in \emptyset} \tau_i\right) \left( \prod_{i \in \sublinks} \left(1 - \tau_i\right) \right) \left(1 - \prod_{i \in \emptyset} \alpha_i\right)\\
&= \sum_{M \in \mathcal{P}\left(\sublinks\right)} \left(\prod_{i \in M} \tau_i\right) \left( \prod_{i \in \sublinks \setminus M} \left(1 - \tau_i\right) \right) \left(1 - \prod_{i \in M} \alpha_i\right) - 0\\
&= \sum_{M \in \mathcal{P}\left(\sublinks\right)} \left(\prod_{i \in M} \tau_i\right) \left( \prod_{i \in \sublinks \setminus M} \left(1 - \tau_i\right) \right)
- \sum_{M \in \mathcal{P}\left(\sublinks\right)} \left(\prod_{i \in M} \tau_i\right) \left( \prod_{i \in \sublinks \setminus M} \left(1 - \tau_i\right) \right) \prod_{i \in M} \alpha_i\\
&= \sum_{M \in \mathcal{P}\left(\sublinks\right)} \left(\prod_{i \in M} \tau_i\right) \left( \prod_{i \in \sublinks \setminus M} \left(1 - \tau_i\right) \right)
- \sum_{M \in \mathcal{P}\left(\sublinks\right)} \left(\prod_{i \in M} \tau_i\alpha_i\right) \left( \prod_{i \in \sublinks \setminus M} \left(1 - \tau_i\right) \right). \label{math:alphapktsplit}
\end{align}

Note that for $j \in \sublinks$ holds
\begin{align}
\mathcal{P}\left(\sublinks\right) = \{M \cup \{j\}\,|\,M \in \mathcal{P}\left(\sublinks \setminus \{j\}\right)\} \cup \{M\,|\,M \in \mathcal{P}\left(\sublinks \setminus \{j\}\right)\}.
\end{align}

Therefore, the second term can be split up
\begin{align}
&\sum_{M \in \mathcal{P}\left(\sublinks\right)} \left(\prod_{i \in M} \tau_i\alpha_i\right) \left( \prod_{i \in \sublinks \setminus M} \left(1 - \tau_i\right) \right)\\
&= \sum_{M \in \mathcal{P}\left(\sublinks \setminus \{j\} \right)} \left(\prod_{i \in (M \cup \{j\})} \tau_i\alpha_i\right) \left( \prod_{i \in (\sublinks) \setminus (M \cup \{j\}) } \left(1 - \tau_i\right) \right)
	+ \sum_{M \in \mathcal{P}\left(\sublinks \setminus \{j\} \right)} \left(\prod_{i \in M} \tau_i\alpha_i\right) \left( \prod_{i \in \sublinks \setminus M} \left(1 - \tau_i\right) \right)\\
	&= \tau_j\alpha_j \sum_{M \in \mathcal{P}\left(\sublinks \setminus \{j\} \right)} \left(\prod_{i \in M} \tau_i\alpha_i\right) \left( \prod_{i \in (\sublinks \setminus \{j\}) \setminus M} \left(1 - \tau_i\right) \right) + 
	(1- \tau_n) \sum_{M \in \mathcal{P}\left(\sublinks \setminus \{j\} \right)} \left(\prod_{i \in M} \tau_i\alpha_i\right) \left( \prod_{i \in (\sublinks \setminus \{j\}) \setminus M} \left(1 - \tau_i\right) \right)\\
	&= (\tau_j\alpha_j + (1-\tau_j)) \sum_{M \in \mathcal{P}\left(\sublinks \setminus \{j\} \right)} \left(\prod_{i \in M} \tau_i\alpha_i\right) \left( \prod_{i \in (\sublinks \setminus \{j\}) \setminus M} \left(1 - \tau_i\right) \right).
\end{align}
This step can be recursively repeated for all nodes in $\sublinks$, finally resulting in
\begin{align}
	\sum_{M \in \mathcal{P}\left(\sublinks\right)} \left(\prod_{i \in M} \tau_i\alpha_i\right) \left( \prod_{i \in \sublinks \setminus M} \left(1 - \tau_i\right) \right) &= \prod_{l \in \sublinks} (\tau_j\alpha_j + (1-\tau_j)).
\intertext{The same calculation for the first term in (\ref{math:alphapktsplit}) results in}
\sum_{M \in \mathcal{P}\left(\sublinks\right)} \left(\prod_{i \in M} \tau_i\right) \left( \prod_{i \in \sublinks \setminus M} \left(1 - \tau_i\right) \right) &= \prod_{j \in \sublinks} (\tau_j + (1-\tau_j)) = 1.
\intertext{Finally, the result is}
\somesending{\sublinks} &= 1 - \prod_{j \in \sublinks} (\tau_j\alpha_j + (1-\tau_j)).
\end{align}

\end{document}